\documentclass[
superscriptaddress,
 amsmath,amssymb,
 aps,
prl,
twocolumn,
]{revtex4-2}
\newcommand{\barium}{$^{133}$Ba$^+$}
\usepackage{braket}

\usepackage{graphicx}

\usepackage[colorlinks=true, allcolors=blue]{hyperref}

\usepackage{upgreek}
\newcommand{\shrinkify}[1]{\textstyle {#1} \displaystyle}

\begin{document}

\title{Errors in stimulated-Raman-induced logic gates in $^{133}$Ba$^+$}

\author{Matthew J. Boguslawski}
\thanks{equal contribution}
\affiliation{Department of Physics and Astronomy, University of California Los Angeles, Los Angeles, CA, USA}
\affiliation{Challenge Institute for Quantum Computation, University of California Los Angeles, Los Angeles, CA, USA}
\email{mbog@physics.ucla.edu}
\email{zwall@physics.ucla.edu}
\author{Zachary J. Wall}
\thanks{equal contribution}
\affiliation{Department of Physics and Astronomy, University of California Los Angeles, Los Angeles, CA, USA}
\author{Samuel R. Vizvary}
\affiliation{Department of Physics and Astronomy, University of California Los Angeles, Los Angeles, CA, USA}
\author{Isam Daniel Moore}
\affiliation{Department of Physics, University of Oregon, Eugene, OR, USA}
\author{Michael Bareian}
\affiliation{Department of Physics and Astronomy, University of California Los Angeles, Los Angeles, CA, USA}
\author{David T. C. Allcock}
\affiliation{Department of Physics, University of Oregon, Eugene, OR, USA}
\author{David J. Wineland}
\affiliation{Department of Physics, University of Oregon, Eugene, OR, USA}
\author{Eric R. Hudson}
\affiliation{Department of Physics and Astronomy, University of California Los Angeles, Los Angeles, CA, USA}
\affiliation{Challenge Institute for Quantum Computation, University of California Los Angeles, Los Angeles, CA, USA}
\affiliation{Center for Quantum Science and Engineering, University of California Los Angeles, Los Angeles, CA, USA}
\author{Wesley C. Campbell}
\affiliation{Department of Physics and Astronomy, University of California Los Angeles, Los Angeles, CA, USA}
\affiliation{Challenge Institute for Quantum Computation, University of California Los Angeles, Los Angeles, CA, USA}
\affiliation{Center for Quantum Science and Engineering, University of California Los Angeles, Los Angeles, CA, USA}

\date{\today}

\begin{abstract}
${}^{133}\mathrm{Ba}^+$ is illuminated by a laser that is far-detuned from optical transitions, and the resulting spontaneous Raman scattering rate is measured. The observed scattering rate is lower than previous theoretical estimates. The majority of the discrepancy is explained by a more accurate treatment of the scattered photon density of states. This work establishes that, contrary to previous models, there is no fundamental limit to laser-driven quantum gates from laser-induced spontaneous Raman scattering. 
\end{abstract}

\maketitle

Trapped ion quantum systems have realized the highest-fidelity single- and two-qubit gate operations to date \cite{Ballance16,Gaebler2016,Bruzewicz19,Christensen20,GTRI2021}. 
Nonetheless, these fidelities remain below estimated thresholds for realizing practical, fault-tolerant quantum computers~\cite{knill05}.
As technical limitations on fidelity are overcome, it becomes increasingly important to understand any \emph{fundamental} limitations on gate fidelity, which is to say those that are inherent to a particular architecture.

For the case of stimulated-Raman laser-driven gates on trapped ion qubits, spontaneous Raman scattering during gates is a chief source of infidelity. 
This process occurs when a far-off-resonant laser photon is incident and another photon is spontaneously emitted, resulting in the ion either returning to its initial state (spontaneous Rayleigh scattering) or transitioning to a different state (spontaneous Raman scattering, SRS)~\cite{Plenio1997Decoherence,Wineland1998Experimental,Wineland2003Quantum,Ozeri07,Uys2010,KenBrown2021}.
In SRS, the spontaneously emitted photon carries information about the ion internal state, thereby limiting the achievable fidelity.

Previous models of this process, here referred to as the constant density of states approximation (CDA) ~\cite{Wineland2003Quantum,Ozeri07} posited the existence of a fundamental lower limit, $\varepsilon_{\mathrm{D}\infty}$, to the achievable gate infidelity due to spontaneous scattering to the metastable ${}^2\mathrm{D}_J$ states present in commonly used species.
However, a recent treatment (referred to here as the ``Moore \textit{et al.}\ treatment'') of photon scattering \cite{Moore22} that includes the frequency-dependence of the density of states predicts no such fundamental limit.

\begin{figure}[t]
  \centering
  \includegraphics[width=\columnwidth]{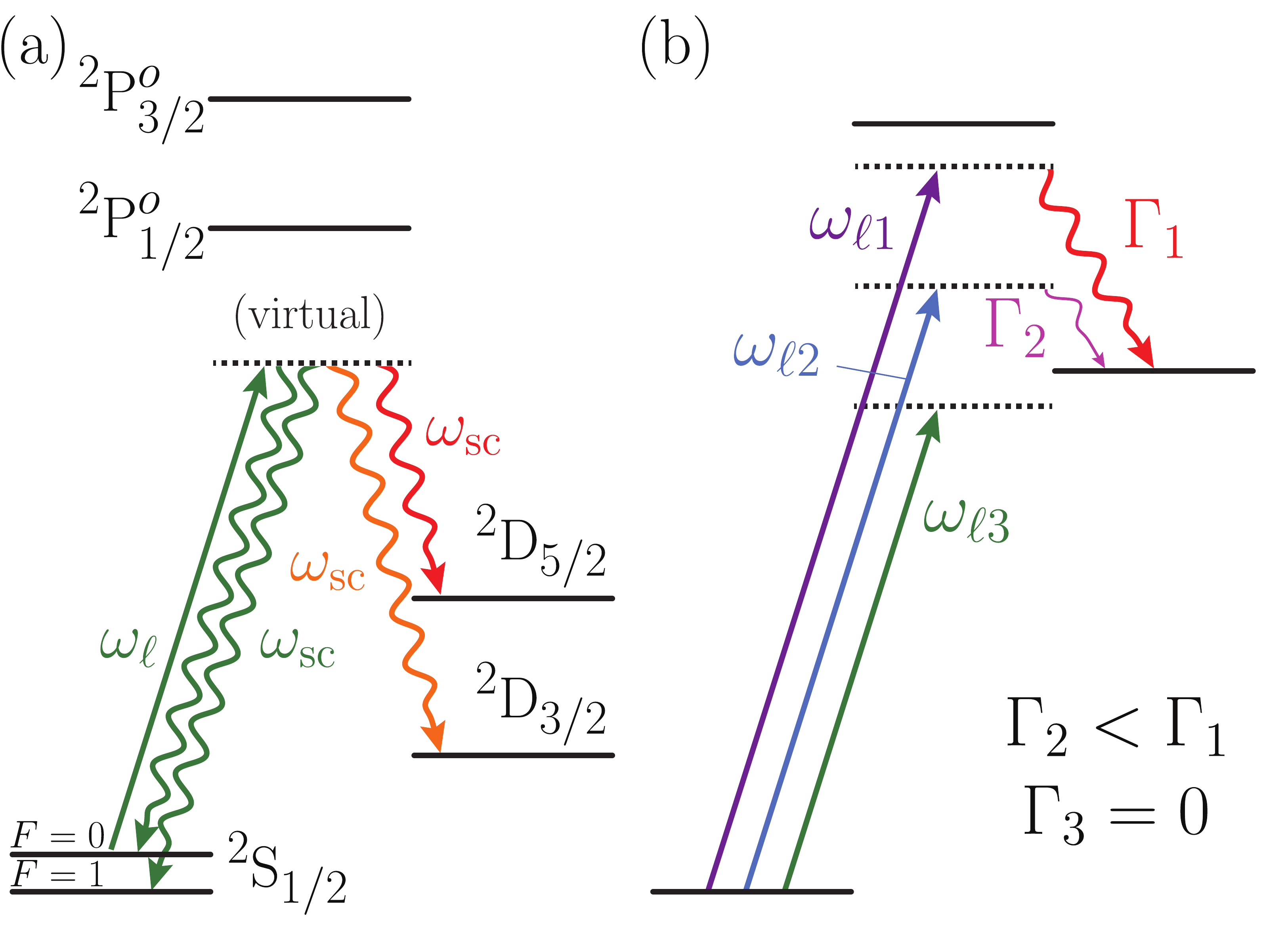}
   \caption{(a) ${}^{133}\mathrm{Ba}^+$ level structure. The SRS photon frequency, $\omega_\mathrm{sc}$, varies depending on the laser frequency, $\omega_\ell$, and the decay channel (wavy lines). (b) As the gate laser is detuned further to the red of the intermediate resonance ($\omega_{\ell i}$, $i = 1 \rightarrow 3$), the scattering rate to a metastable state ($\Gamma_i$, proportional to  $\omega^3_\mathrm{sc}$) decreases until the error channel is closed (see $\omega_\mathrm{sc}^3 \Theta(\omega_\mathrm{sc})$ in Eqn.~\ref{eqn:Gamma2}).}\label{fig:scatter_detuning}
  \vspace{-10pt}
\end{figure}

In this Letter, we measure spontaneous Raman scattering of a trapped \barium\ ion qubit under laser illumination (Fig.~\ref{fig:scatter_detuning}).
Our results disagree with CDA models of the scattering process~\cite{Wineland2003Quantum,Ozeri07}, but agree with the Moore \textit{et al.}\ treatment~\cite{Moore22}. 
In what follows, we detail a model of the scattering process that reproduces the main features of the experiment, describe the experimental approach and results, and discuss implications for trapped ion quantum information processing. 

The salient features of the observed scattering rate can be understood with a model that stops short of the Moore \textit{et al.}\ treatment of Ref.~\cite{Moore22}. Using second-order perturbation theory, the scattering rate from initial state $|i\rangle$ (with energy $\hbar \omega_i)$ to final state $|f\rangle$ ($\hbar \omega_f$) through the intermediate states $|k\rangle$ ($\hbar \omega_k)$ 
can be estimated as:
\begin{equation}\label{eqn:Gamma2}
    \Gamma_{i\rightarrow f} = \frac{\mathcal{E}_0^2 }{4\hbar^2}\frac{\omega_\mathrm{sc}^3 \Theta(\omega_\mathrm{sc})}{3 \pi \epsilon_0 \hbar c^3}\sum_q \left| \sum_k \frac{\langle f | d_q | k \rangle \langle k | d_\ell |i\rangle}{\omega_{ik} - \omega_\ell}\right|^2.
\end{equation}
Here, $\Theta(x)$ is the Heaviside step function; $\omega_\mathrm{sc} \equiv \omega_\ell-\omega_{if}$ is the scattered photon frequency, given by the difference between the frequency of the incident laser photon, $\omega_\ell$, and $\omega_{if} \equiv \omega_f - \omega_i$; $\omega_{ik} \equiv \omega_k - \omega_i$; $d_\ell \equiv \mathbf{d} \cdot \mbox{\boldmath$\hat{\epsilon}$}_\ell$ and $d_q \equiv \mathbf{d} \cdot \mathbf{\hat{e}}_q$ are components of the electric dipole operator $\mathbf{d} \equiv \sum_j e \mathbf{r}_j$.  The role of $\Theta (\omega_\mathrm{sc})$ is to enforce energy conservation when the laser is too red to populate a particular potential final state via single-photon absorption (see Fig.~\ref{fig:scatter_detuning}(b)).
The applied laser, which we model as monochromatic, produces an electric field at the ion of 
$\mathbf{E}(t) = \frac{\mathcal{E}_o}{2} (
\mbox{\boldmath$\hat{\epsilon}$}_\ell\, e^{-i \omega_\ell t} 
+ \mbox{\boldmath$\hat{\epsilon}$}^\ast_\ell e^{i \omega_\ell t}
)$.
The total spontaneous Raman scattering rate is calculated by summing Eqn.~\ref{eqn:Gamma2} over all final states $| f \rangle$ with $f \neq i$ (i.e. omits Rayleigh scattering). 
Though this model, which we refer to as the $\omega^3$-theory model, neglects emission-first processes, and we restrict our basis to the five lowest electronic states, it reproduces the observed scattering rate at the ten percent level. 
Explicit expressions for the SRS rates in the CDA and $\omega^3$-theory models in terms of Einstein $A$ coefficients are given in the Supplemental Information \cite{SI}.

To empirically compare the scattering behavior of trapped ions in the far-detuned regime to these theories, we illuminate a trapped ion in a single initial quantum state with far-detuned light and probe for internal state changes induced due to SRS.  
Since far-off-resonance scattering is rare, high-fidelity state preparation and measurement (SPAM) is desirable to discern scattering from SPAM errors, and we therefore perform this measurement with $^{133}\text{Ba}^+$ \cite{Christensen20}. 
Specifically, a single $^{133}\text{Ba}^+$ ion (see Fig.~\ref{fig:scatter_detuning}(a))  is confined in a linear Paul trap with a minimum ion-electrode spacing of 3~mm driven at $\Omega = 2\pi \times 2.6$~MHz, resulting in a radial secular frequency of $\omega_r = 2\pi \times 230$~kHz. 
Ions are detected by imaging the laser induced fluorescence (LIF) of the $^2\text{S}_{1/2}$ to $^2\text{P}^o_{1/2}$ transition at 493~nm through an objective with a numerical aperture of $0.28$. 
The ion is illuminated by a Continuum Verdi-V10, 532~nm laser focused to a $1/e^2$ intensity radius $w_0\approx40\,\,\upmu \mathrm{m}$ centered on the ion with optical power between 0.3~W and 1.4~W. 
This wavelength is a suitable choice for driving gates in both $^2\mathrm{S}_{1/2}$ $g$-type qubits and $^2\mathrm{D}_{5/2}$ $m$-type qubits in $^{133}\text{Ba}^+$ \cite{Allcock2021omg}.
The polarization of the light is set to drive $\sigma^+$ transitions, with a 0.5~mT magnetic field at the ion aligned antiparallel to the laser beam $\mathbf{k}$ vector \footnote{Beam polarization and bias field alignment was determined using measurements of polarization-dependent Stark shifts on the ion. Beam polarization was left-hand circular with 99.86\% degree of circular polarization at the vacuum viewport, measured with a Thorlabs PAX1000VIS polarimeter. The bias field was aligned to within an estimated 2$^{\circ}$ from the $\mathbf{k}$ vector axis.} 
The laser intensity is determined from a differential AC Stark shift measurement of the $o$-type clock-state qubit defined on $|^2\text{S}_{1/2},F=1,m_F=0\rangle$ and $|^2\text{D}_{5/2},F=3,m_F=0\rangle$ via narrow-linewidth optical spectroscopy at 1762~nm.

The total SRS rate measurement proceeds by first preparing the ion in the $\ket{\uparrow} \equiv \ket{F=0, m_F=0}$ $g$-type clock qubit state of the $^2\mathrm{S}_{1/2}$ manifold via optical pumping~\cite{Christensen20}. 
The ion is then illuminated by the 532~nm laser for a chosen exposure time between 4~ms  and 11~ms, during which the ion may spontaneously scatter a photon (see Fig.~\ref{fig:scatter_detuning}(a)).
Next,  a $300\,\,\upmu \mathrm{s}$ pulse of resonant 455~nm, 585~nm and 650~nm light is used to optically pump (``shelve'') the ion if it is in the  $^2\mathrm{S}_{1/2}(F=1)$ or $^2\mathrm{D}_{3/2}$ manifolds to $^2\mathrm{D}_{5/2}$ with high fidelity \cite{Christensen20}. 
Finally, population in $\ket{\uparrow}$ is measured by monitoring LIF while Doppler cooling via 493~nm and 650~nm illumination for~8 ms, and a measurement that does not yield LIF indicates that a SRS event occurred. 
The fraction of such ``dark'' events recorded in the same sequence without the 532~nm laser applied is subtracted from the laser-on experiments to yield a background-corrected probability of SRS events. 
Given the small probability of a SRS event, the experiment is repeated for $>5\cdot10^4$ measurements to collect sufficient statistics, and the probability of multiple SRS events in a single illumination time is negligible.
The SRS rate to only the $^2$D$_{5/2}$ manifold is measured in the same manner, omitting the shelving step. 
For the total SRS rate measurement, the observed background event probability was $6(2)\cdot10^{-4}$, primarily due to shelving errors.
As the measurement of scattering to $^2$D$_{5/2}$ requires no shelving step, background events are due primarily to state preparation errors, which occur with probability $4(4)\cdot10^{-5}$. 

\begin{figure}[t]
\includegraphics[width=0.48\textwidth]{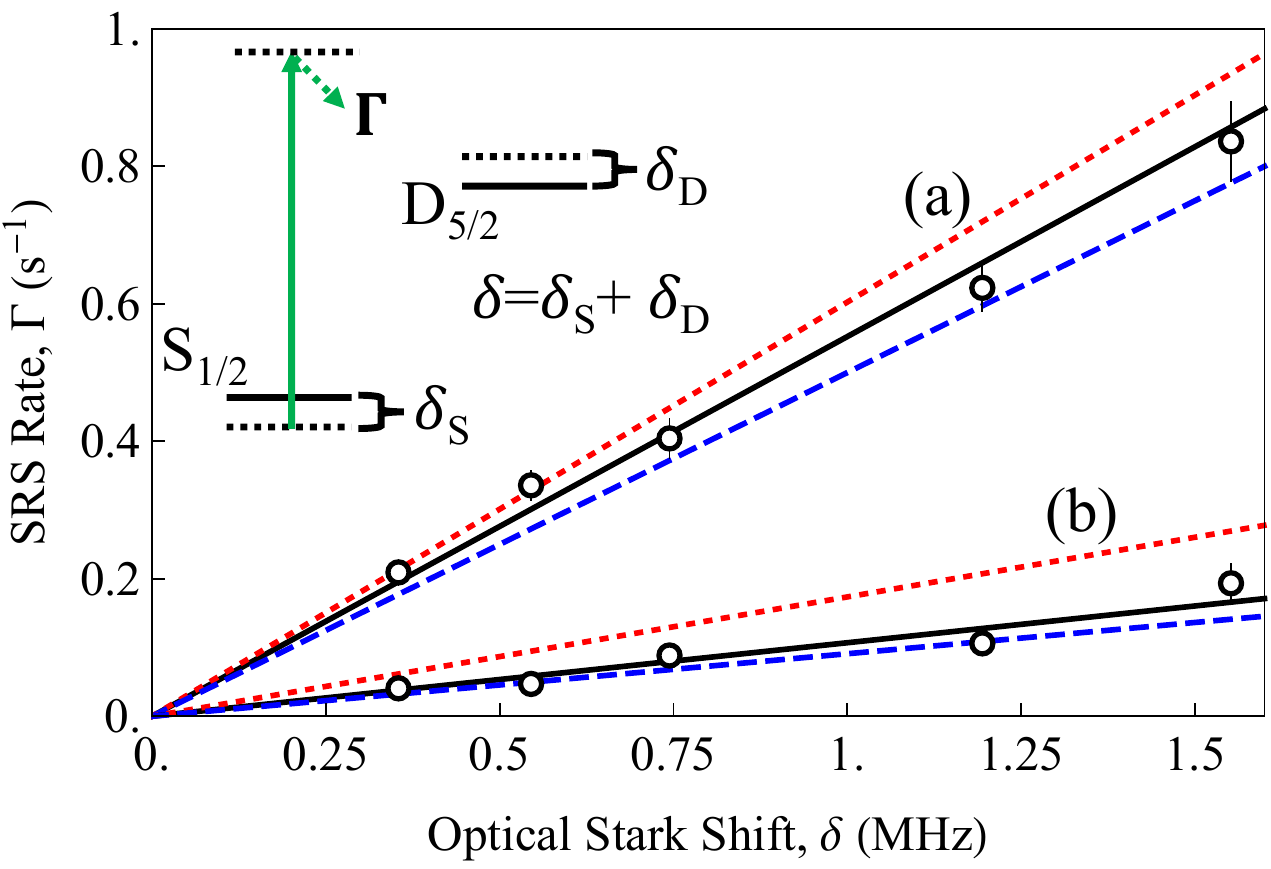}
\caption{\label{fig:GammaSponTotal} SRS rate, $\Gamma$, from 532~nm laser illumination (green) versus differential $o$-type Stark shift, $\delta$. The upper grouping (a) shows the total SRS rate, while the lower grouping (b) shows rate for SRS into only $^2$D$_{5/2}$. (Black circles) Measurements of the scattering rate. (Red, dotted) CDA theory prediction. (Black, solid) Moore \textit{et al.}\ treatment \cite{Moore22}. (Blue. dashed) $\omega^3$-theory prediction. 
}
\end{figure}

Figure~\ref{fig:GammaSponTotal}(a) shows the measured total SRS rate plotted against the differential laser AC Stark shift of the $o$-type qubit and Fig.~\ref{fig:GammaSponTotal}(b) shows the measured the SRS rate into only the $^2\mathrm{D}_{5/2}$ manifold.
The scattering rate $\Gamma_\mathrm{meas}$ is extracted from the laser exposure time, $\tau_\ell$, and the background-corrected, measured probability of scattering, $P_\mathrm{meas}$, as $\Gamma_\mathrm{meas} = -\ln{\left(1 - P_\mathrm{meas}\right)}/\tau_\ell$.
The vertical error bars are calculated from the Wilson score interval for the measured scattering events.
The horizontal uncertainties are the standard errors of the spectral peak center of the optical spectra with and without the laser-induced Stark shift added in quadrature (smaller than the width of the data markers in Fig.~\ref{fig:GammaSponTotal}).

A linear best fit is calculated using an orthogonal distance regression (ODR)~\cite{boggs1992user}, taking into account statistical uncertainty in the measurements of the scattering rate and the Stark shift frequency. The Stark shift is calculated as a function of intensity and used to convert to laser intensity \cite{Wineland2003Quantum}. The results are summarized in Table~\ref{tab:GammaSlopes}.
The data show good agreement with the Moore \textit{et al.}\ treatment of spontaneous Raman scattering \cite{Moore22}. 
The $\omega^3$-theory prediction for the total SRS rate shows a $10\%$ offset due to the neglect of the emission-first terms that are present in the Moore \textit{et al.}\ treatment. 
The measurements are in disagreement with the prediction of the CDA model, particularly for the SRS rate to ${}^2\mathrm{D}_{5/2}$ states; as this process produces a lower energy photon $\hbar \omega_\mathrm{sc}$ than scattering to a ${}^2\mathrm{S}_{1/2}$ state, leading to the observed larger deviation of the CDA from the data for the ${}^2\mathrm{D}_{5/2}$ SRS process. 

\begin{table}[t]
\caption{\label{tab:GammaSlopes}%
Comparison of models with the measured value of SRS rate vs.\ laser intensity. Values are reported in units of SRS rate ($10^{-9}$) s$^{-1}$~per~(W/m$^2$) intensity.}
\begin{ruledtabular}
\begin{tabular}{ccccc}
SRS Type& CDA \cite{Ozeri07}& $\omega^3$ \cite{SI} &Moore \textit{et al.}\ \cite{Moore22}&Measured\\
\hline
Total ($\Gamma_\mathrm{SRS}$)&1.65&1.37&1.52& $\mathbf{1.52(5)}$\\
$\mathrm{D}_{5/2} \,\, ( \Gamma_{\mathrm{D}_{5/2}})$& 0.48 & 0.25 & 0.29 & $\mathbf{0.28(2)}$ \\
\end{tabular}
\end{ruledtabular}
\end{table}

\begin{figure}[t]
\includegraphics[width=.48\textwidth]{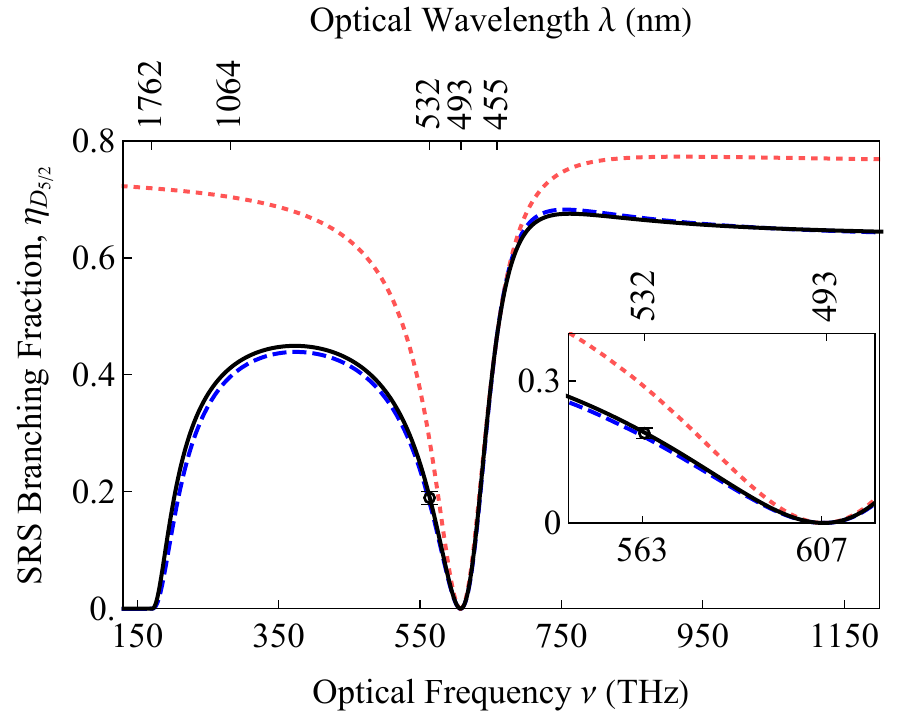}
\caption{\label{fig:BranchingRatiofigure} The SRS branching fraction to the $\mathrm{D}_{5/2}$ manifold, $\eta_{\mathrm{D}_{5/2}}$ as a function of laser frequency.
(Black, solid) Moore \textit{et al.}\ treatment \cite{Moore22}. (Red, dashed) CDA model. (Blue, dashed) $\omega^3$-theory prediction.
The black point shows the result of 21 measurements with the standard error displayed as error bars. The inset shows a zoom-in  near the gate-laser frequency.}
\end{figure}

To explore this disagreement further, the SRS branching \emph{fraction} to the ${}^2\mathrm{D}_{5/2}$ manifold, $\eta_{\mathrm{D}_{5/2}} \equiv \Gamma_{\mathrm{D}_{5/2}}/\Gamma_\mathrm{SRS}$ was separately measured, where $\Gamma_{\mathrm{D}_{5/2}}$ and $\Gamma_\mathrm{SRS}$ are the ${}^2\mathrm{D}_{5/2}$ and total SRS rates. 
Although the Stark shift of the laser was used to monitor the intensity stability on the ion, the result is independent of the intensity on the ion and is simply a function of the gate-laser wavelength. Figure~\ref{fig:BranchingRatiofigure} illustrates the striking consequence of the inclusion of the density of states $\omega_\mathrm{sc}^3$ factor (black trace). The CDA model (red dashed trace) shows a large asymptotic value of $\mathrm{lim}_{\omega_\ell\rightarrow 0}\, \eta_{\mathrm{D}_{5/2}}=0.73$, which implies nearly-static fields would show a significant branching fraction to ${}^2\mathrm{D}_{5/2}$, a consequence that would require violation of energy conservation. In contrast, the $\omega^3$-theory model (Eqn.~\ref{eqn:Gamma2}) shows scatter to ${}^2\mathrm{D}_{J}$ smoothly becomes  forbidden once the laser frequency is less than the ${}^2\mathrm{D}_{J}$ transition frequency. 
The measured value of the SRS branching fraction $\eta_{\mathrm{D}_{5/2}}=0.19(1)$ is reproduced by the $\omega^3$-theory model (0.7$\sigma_E$), but is incompatible with the CDA model (8.7$\sigma_E$) --- here $\sigma_E$ is the standard error.

\par
For quantum information applications, it is essential to understand the fundamental limit posed by the far-detuned SRS events. 
For a stimulated-Raman transition, the probability of an ion to undergo an SRS event during the gate time $\tau_\ell$ for gate lasers indexed by $j$ is approximated by 
$ \epsilon \approx 1 - e^{-\sum_j \tau_\ell\Gamma_{j}}$.
For a typical single-qubit gate (modelled here as a $\pi$ pulse) with two Raman beams, the gate time $\tau_{1q} = \pi/(2|\Omega_\mathrm{R}|)$, where $\Omega_\mathrm{R}$ is the resonant, stimulated-Raman Rabi frequency.
For a two-qubit gate that traverses $K$ loops in phase space, the gate time is reduced by the coupling strength to the motional sideband defined by the Lamb-Dicke parameter $\eta$, giving $\tau_{2q} = \pi\sqrt{K}/(2\eta|\Omega_\mathrm{R}'|)$ \cite{SorensenMolmer2000}.
For two-qubit gates, three beams may be used to produce two stimulated-Raman couplings, with the most efficient coupling realized in a counter-propagating configuration with two beams in one direction and the third, with twice the intensity in the opposite direction, increasing the Rabi rate $\Omega_\mathrm{R}'$ by $\sqrt{2}$ compared to $\Omega_R$ for four balanced beams. 
The scattering rates $\Gamma_j$ are given by Eqn.~\ref{eqn:Gamma2}, leading to single- and two-qubit gate errors of, respectively:
\begin{equation}
\begin{split}\label{eqn:1-2-qerrors}
    \epsilon_{1q} &\approx 2\tau_{1q} \Gamma=  \pi \Gamma/|\Omega_\mathrm{R}|\\
    \epsilon_{2q} &\approx (2)3\tau_{2q} \tilde{\Gamma}= 3\pi\sqrt{K}\tilde{\Gamma}/(\eta|\Omega_\mathrm{R}'|).
\end{split}
\end{equation}
Here, the expression for $\epsilon_{2q}$ accounts for the fact that SRS by either ion leads to an error, and $\tilde{\Gamma}$ is the scattering rate from the average beam intensity. 

\begin{figure}[t]
\includegraphics[width=.48\textwidth]{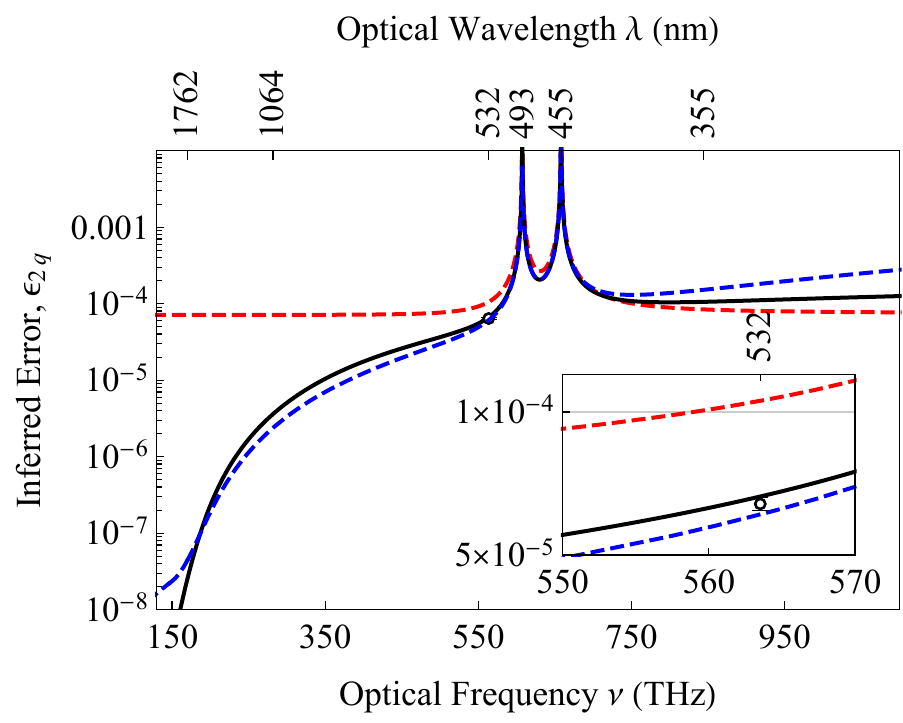}
\caption{\label{fig:GateError} Calculated two-qubit gate error versus gate-laser optical frequency. (Black, solid) Moore \textit{et al.}\ theory using a three-beam configuration, $K=1$ and motional sideband frequency of $2\pi\times5$~MHz. (Red, dashed) CDA theory for three beams, which reproduces limit $\varepsilon_{\mathrm{D}\infty} =(1/2) 1.46\cdot10^{-4}$ for this configuration. (Blue, dashed) $\omega^3$-theory prediction. (Black point) Inferred achievable two-qubit gate error, calculated from the measured scattering rate and Stark shifts at 532~nm. 
}
\end{figure}

\par
Since the stimulated-Raman Rabi frequency $\Omega_\mathrm{R}$ is proportional to the differential AC Stark shift, the fit results from Fig.~\ref{fig:GammaSponTotal} can be used to calculate an empirical value for the achievable error probability for a two-qubit gate at 532~nm
\footnote{The slope of the ODR of the total SRS data in Fig.~\ref{fig:GateError} gives $\Gamma/\delta$ for SRS rate $\Gamma$ and differential Stark shift $\delta$. At 532~nm, $\delta/\Omega_\mathrm{R}$ is calculated following Eqn. 2.3 and 2.10 of \cite{Wineland2003Quantum}. 
Using this quantity, $\pi (\Gamma/\delta)(\delta/\Omega_\mathrm{R})(3\sqrt{K}/\eta) = (\Gamma/\Omega_\mathrm{R})(3\pi\sqrt{K}/\eta) = \epsilon_{2q}$ from Eq.~\ref{eqn:1-2-qerrors} is calculated at 532~nm.}.
In Fig.~\ref{fig:GateError}, we consider a two-qubit gate with $K=1$ operating on a $2\pi \times 5$~MHz motional normal mode and find that the inferred two-qubit gate error is 6.4(2)$\cdot10^{-5}$, in agreement with the Moore \textit{et al.}\ theory value of 6.6$\cdot10^{-5}$ \cite{Moore22}.
This is in contrast to the value provided in Table~III of Ref.~\cite{Ozeri07}, where the CDA results in the conclusion that errors below $\varepsilon_{\mathrm{D}\infty}=(1/2)1.46\cdot10^{-4} = 7.3\cdot10^{-5}$ (for a three beam configuration) detuned outside the ${}^2\mathrm{P}_{J}^o$ levels are not possible in $\mathrm{Ba}^+$ under \emph{any} conditions. Further, for 532~nm, the CDA theory predicts an error of $1.05\cdot10^{-4}$, also in disagreement with our measurement-based estimate.
Notably, the updated models as well as the infidelity inferred from measurements with a 532~nm gate laser reach below the $10^{-4}$ threshold anticipated for efficient error correction \cite{knill05}. 
With the removal of the asymptotic error rate predicted by the CDA model, additional reductions to the SRS error rate can be found by, for example, choosing a gate laser further red-detuned from the ${}^2\mathrm{P}_{J}^o$ levels.

\par
In summary, the Raman scattering rate of ${}^{133}\mathrm{Ba}^+$ under illumination by laser light at 532~nm was studied and found to be smaller than previously predicted. The observed scattering rate agrees with more recent and more complete models, with the difference in models largely due to the inclusion of the scattered photon density of states. This result has important consequences. For example, in contradiction to previous predictions for laser-based gates, there is no fundamental limit to achievable gate error from spontaneous Raman scattering for trapped ion quantum processing since $\epsilon_{2q}\rightarrow0$ as $\omega_\ell\rightarrow0$. Also, in contradiction to previous predictions \cite{Ozeri07}, gate errors below $10^{-4}$ are achievable in ${}^{133}\mathrm{Ba}^+$ at the technologically convenient wavelength of 532~nm. 

\par
We thank Clayton Ho for assistance in calculating Stark shifts. 
This work was supported by the ARO under Grants  No.\ W911NF-20-1-0037 and No.\ W911NF-19-1-0297 and the NSF under Grants No.\ PHY-2110421, No.\ PHY-2207985, and No.\ OMA-2016245.

\bibliographystyle{apsrev4-1}
\bibliography{References}

\providecommand{\noopsort}[1]{}\providecommand{\singleletter}[1]{#1}%
\begin{thebibliography}{19}%
\makeatletter
\providecommand \@ifxundefined [1]{%
 \@ifx{#1\undefined}
}%
\providecommand \@ifnum [1]{%
 \ifnum #1\expandafter \@firstoftwo
 \else \expandafter \@secondoftwo
 \fi
}%
\providecommand \@ifx [1]{%
 \ifx #1\expandafter \@firstoftwo
 \else \expandafter \@secondoftwo
 \fi
}%
\providecommand \natexlab [1]{#1}%
\providecommand \enquote  [1]{``#1''}%
\providecommand \bibnamefont  [1]{#1}%
\providecommand \bibfnamefont [1]{#1}%
\providecommand \citenamefont [1]{#1}%
\providecommand \href@noop [0]{\@secondoftwo}%
\providecommand \href [0]{\begingroup \@sanitize@url \@href}%
\providecommand \@href[1]{\@@startlink{#1}\@@href}%
\providecommand \@@href[1]{\endgroup#1\@@endlink}%
\providecommand \@sanitize@url [0]{\catcode `\\12\catcode `\$12\catcode
  `\&12\catcode `\#12\catcode `\^12\catcode `\_12\catcode `\%12\relax}%
\providecommand \@@startlink[1]{}%
\providecommand \@@endlink[0]{}%
\providecommand \url  [0]{\begingroup\@sanitize@url \@url }%
\providecommand \@url [1]{\endgroup\@href {#1}{\urlprefix }}%
\providecommand \urlprefix  [0]{URL }%
\providecommand \Eprint [0]{\href }%
\providecommand \doibase [0]{http://dx.doi.org/}%
\providecommand \selectlanguage [0]{\@gobble}%
\providecommand \bibinfo  [0]{\@secondoftwo}%
\providecommand \bibfield  [0]{\@secondoftwo}%
\providecommand \translation [1]{[#1]}%
\providecommand \BibitemOpen [0]{}%
\providecommand \bibitemStop [0]{}%
\providecommand \bibitemNoStop [0]{.\EOS\space}%
\providecommand \EOS [0]{\spacefactor3000\relax}%
\providecommand \BibitemShut  [1]{\csname bibitem#1\endcsname}%
\let\auto@bib@innerbib\@empty
\bibitem [{\citenamefont {Ballance}\ \emph {et~al.}(2016)\citenamefont
  {Ballance}, \citenamefont {Harty}, \citenamefont {Linke}, \citenamefont
  {Sepiol},\ and\ \citenamefont {Lucas}}]{Ballance16}%
  \BibitemOpen
  \bibfield  {author} {\bibinfo {author} {\bibfnamefont {C.~J.}\ \bibnamefont
  {Ballance}}, \bibinfo {author} {\bibfnamefont {T.~P.}\ \bibnamefont {Harty}},
  \bibinfo {author} {\bibfnamefont {N.~M.}\ \bibnamefont {Linke}}, \bibinfo
  {author} {\bibfnamefont {M.~A.}\ \bibnamefont {Sepiol}}, \ and\ \bibinfo
  {author} {\bibfnamefont {D.~M.}\ \bibnamefont {Lucas}},\ }\href
  {https://doi.org/10.1103/PhysRevLett.117.060504} {\bibfield  {journal}
  {\bibinfo  {journal} {Phys. Rev. Lett.}\ }\textbf {\bibinfo {volume} {117}},\
  \bibinfo {pages} {060504} (\bibinfo {year} {2016})}\BibitemShut {NoStop}%
\bibitem [{\citenamefont {Gaebler}\ \emph {et~al.}(2016)\citenamefont
  {Gaebler}, \citenamefont {Tan}, \citenamefont {Lin}, \citenamefont {Wan},
  \citenamefont {Bowler}, \citenamefont {Keith}, \citenamefont {Glancy},
  \citenamefont {Coakley}, \citenamefont {Knill}, \citenamefont {Leibfried},\
  and\ \citenamefont {Wineland}}]{Gaebler2016}%
  \BibitemOpen
  \bibfield  {author} {\bibinfo {author} {\bibfnamefont {J.~P.}\ \bibnamefont
  {Gaebler}}, \bibinfo {author} {\bibfnamefont {T.~R.}\ \bibnamefont {Tan}},
  \bibinfo {author} {\bibfnamefont {Y.}~\bibnamefont {Lin}}, \bibinfo {author}
  {\bibfnamefont {Y.}~\bibnamefont {Wan}}, \bibinfo {author} {\bibfnamefont
  {R.}~\bibnamefont {Bowler}}, \bibinfo {author} {\bibfnamefont {A.~C.}\
  \bibnamefont {Keith}}, \bibinfo {author} {\bibfnamefont {S.}~\bibnamefont
  {Glancy}}, \bibinfo {author} {\bibfnamefont {K.}~\bibnamefont {Coakley}},
  \bibinfo {author} {\bibfnamefont {E.}~\bibnamefont {Knill}}, \bibinfo
  {author} {\bibfnamefont {D.}~\bibnamefont {Leibfried}}, \ and\ \bibinfo
  {author} {\bibfnamefont {D.~J.}\ \bibnamefont {Wineland}},\ }\href
  {https://doi.org/10.1103/PhysRevLett.117.060505} {\bibfield  {journal}
  {\bibinfo  {journal} {Phys. Rev. Lett.}\ }\textbf {\bibinfo {volume} {117}},\
  \bibinfo {pages} {060505} (\bibinfo {year} {2016})}\BibitemShut {NoStop}%
\bibitem [{\citenamefont {Bruzewicz}\ \emph {et~al.}(2019)\citenamefont
  {Bruzewicz}, \citenamefont {Chiaverini}, \citenamefont {McConnell},\ and\
  \citenamefont {Sage}}]{Bruzewicz19}%
  \BibitemOpen
  \bibfield  {author} {\bibinfo {author} {\bibfnamefont {C.~D.}\ \bibnamefont
  {Bruzewicz}}, \bibinfo {author} {\bibfnamefont {J.}~\bibnamefont
  {Chiaverini}}, \bibinfo {author} {\bibfnamefont {R.}~\bibnamefont
  {McConnell}}, \ and\ \bibinfo {author} {\bibfnamefont {J.~M.}\ \bibnamefont
  {Sage}},\ }\href {\doibase 10.1063/1.5088164} {\bibfield  {journal} {\bibinfo
   {journal} {Appl. Phys. Rev.}\ }\textbf {\bibinfo {volume} {6}},\ \bibinfo
  {pages} {021314} (\bibinfo {year} {2019})}\BibitemShut {NoStop}%
\bibitem [{\citenamefont {Christensen}\ \emph {et~al.}(2020)\citenamefont
  {Christensen}, \citenamefont {Hucul}, \citenamefont {Campbell},\ and\
  \citenamefont {Hudson}}]{Christensen20}%
  \BibitemOpen
  \bibfield  {author} {\bibinfo {author} {\bibfnamefont {J.~E.}\ \bibnamefont
  {Christensen}}, \bibinfo {author} {\bibfnamefont {D.}~\bibnamefont {Hucul}},
  \bibinfo {author} {\bibfnamefont {W.~C.}\ \bibnamefont {Campbell}}, \ and\
  \bibinfo {author} {\bibfnamefont {E.~R.}\ \bibnamefont {Hudson}},\ }\href
  {\doibase 10.1038/s41534-020-0265-5} {\bibfield  {journal} {\bibinfo
  {journal} {npj Quantum Inf}\ }\textbf {\bibinfo {volume} {6}},\ \bibinfo
  {pages} {35} (\bibinfo {year} {2020})}\BibitemShut {NoStop}%
\bibitem [{\citenamefont {Clark}\ \emph {et~al.}(2021)\citenamefont {Clark},
  \citenamefont {Tinkey}, \citenamefont {Sawyer}, \citenamefont {Meier},
  \citenamefont {Burkhardt}, \citenamefont {Seck}, \citenamefont {Shappert},
  \citenamefont {Guise}, \citenamefont {Volin}, \citenamefont {Fallek},
  \citenamefont {Hayden}, \citenamefont {Rellergert},\ and\ \citenamefont
  {Brown}}]{GTRI2021}%
  \BibitemOpen
  \bibfield  {author} {\bibinfo {author} {\bibfnamefont {C.~R.}\ \bibnamefont
  {Clark}}, \bibinfo {author} {\bibfnamefont {H.~N.}\ \bibnamefont {Tinkey}},
  \bibinfo {author} {\bibfnamefont {B.~C.}\ \bibnamefont {Sawyer}}, \bibinfo
  {author} {\bibfnamefont {A.~M.}\ \bibnamefont {Meier}}, \bibinfo {author}
  {\bibfnamefont {K.~A.}\ \bibnamefont {Burkhardt}}, \bibinfo {author}
  {\bibfnamefont {C.~M.}\ \bibnamefont {Seck}}, \bibinfo {author}
  {\bibfnamefont {C.~M.}\ \bibnamefont {Shappert}}, \bibinfo {author}
  {\bibfnamefont {N.~D.}\ \bibnamefont {Guise}}, \bibinfo {author}
  {\bibfnamefont {C.~E.}\ \bibnamefont {Volin}}, \bibinfo {author}
  {\bibfnamefont {S.~D.}\ \bibnamefont {Fallek}}, \bibinfo {author}
  {\bibfnamefont {H.~T.}\ \bibnamefont {Hayden}}, \bibinfo {author}
  {\bibfnamefont {W.~G.}\ \bibnamefont {Rellergert}}, \ and\ \bibinfo {author}
  {\bibfnamefont {K.~R.}\ \bibnamefont {Brown}},\ }\href
  {https://doi.org/10.1103/PhysRevLett.127.130505} {\bibfield  {journal}
  {\bibinfo  {journal} {Phys. Rev. Lett.}\ }\textbf {\bibinfo {volume} {127}},\
  \bibinfo {pages} {130505} (\bibinfo {year} {2021})}\BibitemShut {NoStop}%
\bibitem [{\citenamefont {Knill}(2005)}]{knill05}%
  \BibitemOpen
  \bibfield  {author} {\bibinfo {author} {\bibfnamefont {E.}~\bibnamefont
  {Knill}},\ }\href {\doibase https://doi.org/10.1038/nature03350} {\bibfield
  {journal} {\bibinfo  {journal} {Nature}\ }\textbf {\bibinfo {volume} {434}},\
  \bibinfo {pages} {39} (\bibinfo {year} {2005})}\BibitemShut {NoStop}%
\bibitem [{\citenamefont {Plenio}\ and\ \citenamefont
  {Knight}(1997)}]{Plenio1997Decoherence}%
  \BibitemOpen
  \bibfield  {author} {\bibinfo {author} {\bibfnamefont {M.~B.}\ \bibnamefont
  {Plenio}}\ and\ \bibinfo {author} {\bibfnamefont {P.~L.}\ \bibnamefont
  {Knight}},\ }\href {\doibase 10.1098/rspa.1997.0109} {\bibfield  {journal}
  {\bibinfo  {journal} {Proc. R. Soc. Lond. A.}\ }\textbf {\bibinfo {volume}
  {453}},\ \bibinfo {pages} {2017} (\bibinfo {year} {1997})}\BibitemShut
  {NoStop}%
\bibitem [{\citenamefont {Wineland}\ \emph {et~al.}(1998)\citenamefont
  {Wineland}, \citenamefont {Monroe}, \citenamefont {Itano}, \citenamefont
  {Leibfried}, \citenamefont {King},\ and\ \citenamefont
  {Meekhof}}]{Wineland1998Experimental}%
  \BibitemOpen
  \bibfield  {author} {\bibinfo {author} {\bibfnamefont {D.~J.}\ \bibnamefont
  {Wineland}}, \bibinfo {author} {\bibfnamefont {C.}~\bibnamefont {Monroe}},
  \bibinfo {author} {\bibfnamefont {W.~M.}\ \bibnamefont {Itano}}, \bibinfo
  {author} {\bibfnamefont {D.}~\bibnamefont {Leibfried}}, \bibinfo {author}
  {\bibfnamefont {B.~E.}\ \bibnamefont {King}}, \ and\ \bibinfo {author}
  {\bibfnamefont {D.~M.}\ \bibnamefont {Meekhof}},\ }\href
  {https://tf.nist.gov/general/pdf/1275.pdf} {\bibfield  {journal} {\bibinfo
  {journal} {J. Res. Natl. Inst. Stand. Technol.}\ }\textbf {\bibinfo {volume}
  {103}},\ \bibinfo {pages} {259} (\bibinfo {year} {1998})}\BibitemShut
  {NoStop}%
\bibitem [{\citenamefont {Wineland}\ \emph {et~al.}(2003)\citenamefont
  {Wineland}, \citenamefont {Barrett}, \citenamefont {Britton}, \citenamefont
  {Chiaverini}, \citenamefont {De{M}arco}, \citenamefont {Itano}, \citenamefont
  {Jelenkovi\'c}, \citenamefont {Langer}, \citenamefont {Leibfried},
  \citenamefont {Meyer}, \citenamefont {Rosenband},\ and\ \citenamefont
  {Sch\"atz}}]{Wineland2003Quantum}%
  \BibitemOpen
  \bibfield  {author} {\bibinfo {author} {\bibfnamefont {D.~J.}\ \bibnamefont
  {Wineland}}, \bibinfo {author} {\bibfnamefont {M.}~\bibnamefont {Barrett}},
  \bibinfo {author} {\bibfnamefont {J.}~\bibnamefont {Britton}}, \bibinfo
  {author} {\bibfnamefont {J.}~\bibnamefont {Chiaverini}}, \bibinfo {author}
  {\bibfnamefont {B.}~\bibnamefont {De{M}arco}}, \bibinfo {author}
  {\bibfnamefont {W.~M.}\ \bibnamefont {Itano}}, \bibinfo {author}
  {\bibfnamefont {B.}~\bibnamefont {Jelenkovi\'c}}, \bibinfo {author}
  {\bibfnamefont {C.}~\bibnamefont {Langer}}, \bibinfo {author} {\bibfnamefont
  {D.}~\bibnamefont {Leibfried}}, \bibinfo {author} {\bibfnamefont
  {V.}~\bibnamefont {Meyer}}, \bibinfo {author} {\bibfnamefont
  {T.}~\bibnamefont {Rosenband}}, \ and\ \bibinfo {author} {\bibfnamefont
  {T.}~\bibnamefont {Sch\"atz}},\ }\href {\doibase 10.1098/rsta.2003.1205}
  {\bibfield  {journal} {\bibinfo  {journal} {Phil. Trans. R. Soc. A}\ }\textbf
  {\bibinfo {volume} {361}},\ \bibinfo {pages} {1349} (\bibinfo {year}
  {2003})}\BibitemShut {NoStop}%
\bibitem [{\citenamefont {Ozeri}\ \emph {et~al.}(2007)\citenamefont {Ozeri},
  \citenamefont {Itano}, \citenamefont {R.~B.~Blakestad}, \citenamefont
  {Chiaverini}, \citenamefont {Jost}, \citenamefont {Langer}, \citenamefont
  {Leibfried}, \citenamefont {Reichle}, \citenamefont {Seidelin}, \citenamefont
  {Wesenberg},\ and\ \citenamefont {Wineland}}]{Ozeri07}%
  \BibitemOpen
  \bibfield  {author} {\bibinfo {author} {\bibfnamefont {R.}~\bibnamefont
  {Ozeri}}, \bibinfo {author} {\bibfnamefont {W.~M.}\ \bibnamefont {Itano}},
  \bibinfo {author} {\bibfnamefont {J.~B.}\ \bibnamefont {R.~B.~Blakestad}},
  \bibinfo {author} {\bibfnamefont {J.}~\bibnamefont {Chiaverini}}, \bibinfo
  {author} {\bibfnamefont {J.~D.}\ \bibnamefont {Jost}}, \bibinfo {author}
  {\bibfnamefont {C.}~\bibnamefont {Langer}}, \bibinfo {author} {\bibfnamefont
  {D.}~\bibnamefont {Leibfried}}, \bibinfo {author} {\bibfnamefont
  {R.}~\bibnamefont {Reichle}}, \bibinfo {author} {\bibfnamefont
  {S.}~\bibnamefont {Seidelin}}, \bibinfo {author} {\bibfnamefont {J.~H.}\
  \bibnamefont {Wesenberg}}, \ and\ \bibinfo {author} {\bibfnamefont {D.~J.}\
  \bibnamefont {Wineland}},\ }\href {\doibase 10.1103/PhysRevA.75.042329}
  {\bibfield  {journal} {\bibinfo  {journal} {Phys. Rev. A}\ }\textbf {\bibinfo
  {volume} {75}},\ \bibinfo {pages} {042329} (\bibinfo {year}
  {2007})}\BibitemShut {NoStop}%
\bibitem [{\citenamefont {Uys}\ \emph {et~al.}(2010)\citenamefont {Uys},
  \citenamefont {Biercuk}, \citenamefont {VanDevender}, \citenamefont
  {Ospelkaus}, \citenamefont {Meiser}, \citenamefont {Ozeri},\ and\
  \citenamefont {Bollinger}}]{Uys2010}%
  \BibitemOpen
  \bibfield  {author} {\bibinfo {author} {\bibfnamefont {H.}~\bibnamefont
  {Uys}}, \bibinfo {author} {\bibfnamefont {M.~J.}\ \bibnamefont {Biercuk}},
  \bibinfo {author} {\bibfnamefont {A.~P.}\ \bibnamefont {VanDevender}},
  \bibinfo {author} {\bibfnamefont {C.}~\bibnamefont {Ospelkaus}}, \bibinfo
  {author} {\bibfnamefont {D.}~\bibnamefont {Meiser}}, \bibinfo {author}
  {\bibfnamefont {R.}~\bibnamefont {Ozeri}}, \ and\ \bibinfo {author}
  {\bibfnamefont {J.~J.}\ \bibnamefont {Bollinger}},\ }\href
  {https://doi.org/10.1103/PhysRevLett.105.200401} {\bibfield  {journal}
  {\bibinfo  {journal} {Phys. Rev. Lett.}\ }\textbf {\bibinfo {volume} {105}},\
  \bibinfo {pages} {200401} (\bibinfo {year} {2010})}\BibitemShut {NoStop}%
\bibitem [{\citenamefont {Sawyer}\ and\ \citenamefont
  {Brown}(2021)}]{KenBrown2021}%
  \BibitemOpen
  \bibfield  {author} {\bibinfo {author} {\bibfnamefont {B.~C.}\ \bibnamefont
  {Sawyer}}\ and\ \bibinfo {author} {\bibfnamefont {K.~R.}\ \bibnamefont
  {Brown}},\ }\href {\doibase 10.1103/PhysRevA.103.022427} {\bibfield
  {journal} {\bibinfo  {journal} {Phys. Rev. A}\ }\textbf {\bibinfo {volume}
  {103}},\ \bibinfo {pages} {022427} (\bibinfo {year} {2021})}\BibitemShut
  {NoStop}%
\bibitem [{\citenamefont {Moore}\ \emph {et~al.}(2022)\citenamefont {Moore},
  \citenamefont {Campbell}, \citenamefont {Hudson}, \citenamefont
  {Boguslawski}, \citenamefont {Wineland},\ and\ \citenamefont
  {Allcock}}]{Moore22}%
  \BibitemOpen
  \bibfield  {author} {\bibinfo {author} {\bibfnamefont {I.~D.}\ \bibnamefont
  {Moore}}, \bibinfo {author} {\bibfnamefont {W.}~\bibnamefont {Campbell}},
  \bibinfo {author} {\bibfnamefont {E.~R.}\ \bibnamefont {Hudson}}, \bibinfo
  {author} {\bibfnamefont {M.~J.}\ \bibnamefont {Boguslawski}}, \bibinfo
  {author} {\bibfnamefont {D.~J.}\ \bibnamefont {Wineland}}, \ and\ \bibinfo
  {author} {\bibfnamefont {D.~T.~C.}\ \bibnamefont {Allcock}},\ }\href@noop {}
  {\enquote {\bibinfo {title} {Photon scattering errors during stimulated raman
  transitions in trapped-ion qubits},}\ } (\bibinfo {year} {2022}),\ \Eprint
  {http://arxiv.org/abs/2211.00744} {arXiv:2211.00744 [physics.quant-ph]}
  \BibitemShut {NoStop}%
\bibitem [{SI()}]{SI}%
  \BibitemOpen
  \href@noop {} {}\bibinfo {note} {See Supplemental Material at [URL will be
  inserted by publisher] for explicit formulas for scattering rates in terms of
  Einstein $A$ coefficients.}\BibitemShut {Stop}%
\bibitem [{\citenamefont {Allcock}\ \emph {et~al.}(2021)\citenamefont
  {Allcock}, \citenamefont {Campbell}, \citenamefont {Chiaverini},
  \citenamefont {Chuang}, \citenamefont {Hudson}, \citenamefont {Moore},
  \citenamefont {Ransford}, \citenamefont {Roman}, \citenamefont {Sage},\ and\
  \citenamefont {Wineland}}]{Allcock2021omg}%
  \BibitemOpen
  \bibfield  {author} {\bibinfo {author} {\bibfnamefont {D.~T.~C.}\
  \bibnamefont {Allcock}}, \bibinfo {author} {\bibfnamefont {W.~C.}\
  \bibnamefont {Campbell}}, \bibinfo {author} {\bibfnamefont {J.}~\bibnamefont
  {Chiaverini}}, \bibinfo {author} {\bibfnamefont {I.~L.}\ \bibnamefont
  {Chuang}}, \bibinfo {author} {\bibfnamefont {E.~R.}\ \bibnamefont {Hudson}},
  \bibinfo {author} {\bibfnamefont {I.~D.}\ \bibnamefont {Moore}}, \bibinfo
  {author} {\bibfnamefont {A.}~\bibnamefont {Ransford}}, \bibinfo {author}
  {\bibfnamefont {C.}~\bibnamefont {Roman}}, \bibinfo {author} {\bibfnamefont
  {J.~M.}\ \bibnamefont {Sage}}, \ and\ \bibinfo {author} {\bibfnamefont
  {D.~J.}\ \bibnamefont {Wineland}},\ }\href
  {https://doi.org/10.1063/5.0069544} {\bibfield  {journal} {\bibinfo
  {journal} {Appl. Phys. Lett.}\ }\textbf {\bibinfo {volume} {119}},\ \bibinfo
  {pages} {214002} (\bibinfo {year} {2021})}\BibitemShut {NoStop}%
\bibitem [{Note1()}]{Note1}%
  \BibitemOpen
  \bibinfo {note} {Beam polarization and bias field alignment was determined
  using measurements of polarization-dependent Stark shifts on the ion. Beam
  polarization was left-hand circular with 99.86\% degree of circular
  polarization at the vacuum viewport, measured with a Thorlabs PAX1000VIS
  polarimeter. The bias field was aligned to within an estimated 2$^{\circ }$
  from the $\protect \mathbf {k}$ vector axis.}\BibitemShut {Stop}%
\bibitem [{\citenamefont {Boggs}\ \emph {et~al.}(1992)\citenamefont {Boggs},
  \citenamefont {Byrd}, \citenamefont {Rogers},\ and\ \citenamefont
  {Schnabel}}]{boggs1992user}%
  \BibitemOpen
  \bibfield  {author} {\bibinfo {author} {\bibfnamefont {P.~T.}\ \bibnamefont
  {Boggs}}, \bibinfo {author} {\bibfnamefont {R.~H.}\ \bibnamefont {Byrd}},
  \bibinfo {author} {\bibfnamefont {J.~E.}\ \bibnamefont {Rogers}}, \ and\
  \bibinfo {author} {\bibfnamefont {R.~B.}\ \bibnamefont {Schnabel}},\
  }\href@noop {} {\enquote {\bibinfo {title} {User's reference guide for
  \textsc{odrpack} version 2.01: Software for weighted orthogonal distance
  regression},}\ } (\bibinfo {year} {1992})\BibitemShut {NoStop}%
\bibitem [{\citenamefont {S\o{}rensen}\ and\ \citenamefont
  {M\o{}lmer}(2000)}]{SorensenMolmer2000}%
  \BibitemOpen
  \bibfield  {author} {\bibinfo {author} {\bibfnamefont {A.}~\bibnamefont
  {S\o{}rensen}}\ and\ \bibinfo {author} {\bibfnamefont {K.}~\bibnamefont
  {M\o{}lmer}},\ }\href {\doibase 10.1103/PhysRevA.62.022311} {\bibfield
  {journal} {\bibinfo  {journal} {Phys. Rev. A}\ }\textbf {\bibinfo {volume}
  {62}},\ \bibinfo {pages} {022311} (\bibinfo {year} {2000})}\BibitemShut
  {NoStop}%
\bibitem [{Note2()}]{Note2}%
  \BibitemOpen
  \bibinfo {note} {The slope of the ODR of the total SRS data in Fig.~\ref
  {fig:GateError} gives $\Gamma /\delta $ for SRS rate $\Gamma $ and
  differential Stark shift $\delta $. At 532~nm, $\delta /\Omega _\protect
  \mathrm {R}$ is calculated following Eqn. 2.3 and 2.10 of \cite
  {Wineland2003Quantum}. Using this quantity, $\pi (\Gamma /\delta )(\delta
  /\Omega _\protect \mathrm {R})(3\protect \sqrt {K}/\eta ) = (\Gamma /\Omega
  _\protect \mathrm {R})(3\pi \protect \sqrt {K}/\eta ) = \epsilon _{2q}$ from
  Eq.~\ref {eqn:1-2-qerrors} is calculated at 532~nm.}\BibitemShut {Stop}%
\end{thebibliography}%

\clearpage

\onecolumngrid
\renewcommand{\thefigure}{S\arabic{figure}}
\setcounter{figure}{0}
\renewcommand{\theequation}{S.\arabic{equation}}
\setcounter{equation}{0}
\renewcommand{\thetable}{S.\Roman{table}}
\setcounter{table}{0}
\section{Supplemental Materials}
\section{Scattering models}
Two scattering models considered in this paper (which we denote the ``CDA'' and ``$\omega^3$-theory'' models) are explicitly given below.  To establish a common set of notation that can be used for ions with similar atomic structure, we adopt the notation given in Fig.~\ref{fig:SILevelDiagram} and Table~\ref{tab:Notation}.
The Einstein $A$ coefficients, electron angular momentum quantum numbers, and frequencies will be denoted with subscripts accordingly (\textit{e.g.} $A_{\mathrm{e}',\mathrm{g}''}$, $J_\mathrm{g}$, $\omega_{\mathrm{e},\mathrm{g}'}$, \textit{etc.}).

\subsection{CDA theory model}
First, the constant density of states approximation (CDA) theory model, which we base on the previous theory \cite{Ozeri07}, is most valid when the laser detuning from an intermediate state is small compared to the laser's frequency.  The total spontaneous scattering rate (during a $g$-type, zero-field hyperfine clock-state qubit gate) is given by
\begin{align}
    \Gamma_\mathrm{total, CDA} \equiv & \,\, \left(  A_{\mathrm{e},\mathrm{g}} + A_{\mathrm{e},\mathrm{g}'} \right)\times \frac{g^2}{3} \frac{1}{\Delta^2} + \left(  A_{\mathrm{e}',\mathrm{g}} +  A_{\mathrm{e}',\mathrm{g}'} +  A_{\mathrm{e}',\mathrm{g}''}  \right)\times \frac{g^2}{3} \frac{2}{(\Delta - \omega_\mathrm{FS})^2}
\end{align}
where $\Delta = \omega_\ell - E_{\mathrm{P}1/2}/\hbar$ is the detuning of the laser from the ${}^2\mathrm{P}^o_{1/2} \leftrightarrow {}^2\mathrm{S}_{1/2}$ resonance.
The spontaneous Rayleigh scattering rate is
\begin{align}
        \Gamma_\mathrm{Rayleigh, CDA} \equiv & \,\, A_{\mathrm{e}',\mathrm{g}} \frac{g^2}{3}\left( \frac{3 \Delta^2 - 2 \Delta \omega_\mathrm{FS} + \omega_\mathrm{FS}^2/3}{\Delta^2(\Delta - \omega_\mathrm{FS})^2}  \right),
\end{align}
and the spontaneous Raman scattering rate is is difference between these two, $\Gamma_\mathrm{Raman, CDA} = \Gamma_\mathrm{total, CDA} - \Gamma_\mathrm{Rayleigh, CDA}$.  The branching fraction of the spontaneous Raman scattering rate into ${}^2\mathrm{D}_{5/2}$ is
\begin{align}
    \eta_{\mathrm{D}_{5/2}\mathrm{, CDA}} \equiv & \,\, \frac{\Gamma_{\mathrm{D}_{5/2}\mathrm{, CDA}}}{\Gamma_\mathrm{Raman, CDA}} =  A_{\mathrm{e}',\mathrm{g}''} \times \frac{g^2}{3} \left(  \frac{2}{(\Delta - \omega_\mathrm{FS})^2}\right) /\Gamma_\mathrm{Raman, CDA}.
\end{align}

\subsection{$\omega^3$-theory model}
The $\omega^3$-theory model closely resembles the CDA theory model near resonance, but differs significantly far from resonance.  The difference between the two is simply the addition of the dependence of the spontaneous scattering rate on the frequency of the spontaneously scattered photon.  This necessarily includes a step function to ensure that spontaneous scattering events to excited final states turn off when the laser photons don't have enough energy to populate those states (see Fig.~\ref{fig:scatter_detuning}b).  The total spontaneous scattering rate in this model is given by 
\begin{align}
    \Gamma_{\mathrm{total, }\,\omega^3} = & \,\, \left( \frac{\omega_\ell}{\omega_{\mathrm{e},\mathrm{g}}} \right)^3 A_{\mathrm{e},\mathrm{g}} \times \frac{g^2}{3} \left(  \frac{1}{\Delta^2}\right) \nonumber \\
    & + \Theta (\omega_\ell - \omega_{\mathrm{g},\mathrm{g}'}) \left( \frac{\omega_\ell - \omega_{\mathrm{g},\mathrm{g}'}}{\omega_{\mathrm{e},\mathrm{g}'}} \right)^3 A_{\mathrm{e},\mathrm{g}'} \times \frac{g^2}{3} \left(  \frac{1}{\Delta^2}\right) \nonumber \\
    & + \left( \frac{\omega_\ell}{\omega_{\mathrm{e}',\mathrm{g}}} \right)^3 A_{\mathrm{e}',\mathrm{g}} \times \frac{g^2}{3} \left(  \frac{2}{(\Delta - \omega_\mathrm{FS})^2}\right) \nonumber \\
    & + \Theta (\omega_\ell - \omega_{\mathrm{g},\mathrm{g}'}) \left( \frac{\omega_\ell - \omega_{\mathrm{g},\mathrm{g}'}}{\omega_{\mathrm{e}',\mathrm{g}'}} \right)^3 A_{\mathrm{e}',\mathrm{g}'} \times \frac{g^2}{3} \left(  \frac{2}{(\Delta - \omega_\mathrm{FS})^2}\right) \nonumber \\
    & + \Theta (\omega_\ell - \omega_{\mathrm{g},\mathrm{g}''})\left( \frac{\omega_\ell - \omega_{\mathrm{g},\mathrm{g}''}}{\omega_{\mathrm{e}',\mathrm{g}''}} \right)^3 A_{\mathrm{e}',\mathrm{g}''} \times \frac{g^2}{3} \left(  \frac{2}{(\Delta - \omega_\mathrm{FS})^2}\right) .\label{eq:GtotGeneral}
\end{align}
The spontaneous Rayleigh scattering rate is similar to the CDA theory,
\begin{align}
    \Gamma_{\mathrm{Rayleigh, }\,\omega^3} = &\,\,  \left(  \frac{\omega_\ell}{\omega_{\mathrm{e}',\mathrm{g}}}\right)^3 A_{\mathrm{e}',\mathrm{g}}\, \frac{g^2}{3}\left( \frac{3 \Delta^2 - 2 \Delta \omega_\mathrm{FS} + \omega_\mathrm{FS}^2/3}{\Delta^2(\Delta - \omega_\mathrm{FS})^2}  \right) 
\end{align}
and the spontaneous Raman scattering rate is again given by the difference between the total and Rayleigh rates,
\begin{align}
    \Gamma_{\mathrm{Raman, }\,\omega^3} = & \,\,\Gamma_{\mathrm{total, }\,\omega^3} - \Gamma_{\mathrm{Rayleigh, }\,\omega^3} \nonumber \\
    = & \,\,  \left(  \frac{\omega_\ell}{\omega_{\mathrm{e}',\mathrm{g}}}\right)^3 A_{\mathrm{e}',\mathrm{g}}\, \frac{2g^2}{9} \left( \frac{\omega_\mathrm{FS}}{\Delta(\Delta - \omega_\mathrm{FS})} \right)^2 \nonumber \\
    & +\Theta (\omega_\ell - \omega_{\mathrm{g},\mathrm{g}'}) \left( \frac{\omega_\ell - \omega_{\mathrm{g},\mathrm{g}'}}{\omega_{\mathrm{e},\mathrm{g}'}} \right)^3 A_{\mathrm{e},\mathrm{g}'} \times \frac{g^2}{3} \left(  \frac{1}{\Delta^2}\right) \nonumber \\
    & + \Theta (\omega_\ell - \omega_{\mathrm{g},\mathrm{g}'})\left( \frac{\omega_\ell - \omega_{\mathrm{g},\mathrm{g}'}}{\omega_{\mathrm{e}',\mathrm{g}'}} \right)^3 A_{\mathrm{e}',\mathrm{g}'} \times \frac{g^2}{3} \left(  \frac{2}{(\Delta - \omega_\mathrm{FS})^2}\right) \nonumber \\
    & + \Theta (\omega_\ell - \omega_{\mathrm{g},\mathrm{g}''})\left( \frac{\omega_\ell - \omega_{\mathrm{g},\mathrm{g}''}}{\omega_{\mathrm{e}',\mathrm{g}''}} \right)^3 A_{\mathrm{e}',\mathrm{g}''} \times \frac{g^2}{3} \left(  \frac{2}{(\Delta - \omega_\mathrm{FS})^2}\right)
\end{align}
where we have used $A_{\mathrm{e},\mathrm{g}}/\omega_{\mathrm{e},\mathrm{g}}^3 = A_{\mathrm{e},\mathrm{g}'}/\omega_{\mathrm{e},\mathrm{g}'}^3$.  The branching fraction of the spontaneous Raman scattering to the ${}^2\mathrm{D}_{5/2}$ manifold is 
\begin{align}
    \eta_{\mathrm{D}_{5/2}\mathrm{,}\, \omega^3} = & \,\,  \Theta (\omega_\ell - \omega_{\mathrm{g},\mathrm{g}''})\left( \frac{\omega_\ell - \omega_{\mathrm{g},\mathrm{g}''}}{\omega_{\mathrm{e}',\mathrm{g}''}} \right)^3 A_{\mathrm{e}',\mathrm{g}''} \times \frac{g^2}{3} \left(  \frac{2}{(\Delta - \omega_\mathrm{FS})^2}\right) /\Gamma_{\mathrm{Raman, }\, \omega^3}.
\end{align}

\begin{table}
    \centering
    \begin{tabular}{c|c|c|c}
    \hline \hline
         State & Shorthand&$J$ & Energy ($/\hbar$)\\
        \hline
         ${}^2\mathrm{S}_{1/2}$ & $\mathrm{g}$ & $J_{\mathrm{g}} = \shrinkify{\frac{1}{2}}$ & $\equiv 0$ \\
         ${}^2\mathrm{P}_{1/2}^o$ & $\mathrm{e}$ & $J_{\mathrm{e}} = \shrinkify{\frac{1}{2}}$ & $\omega_{\mathrm{e},\mathrm{g}}$ \\
         ${}^2\mathrm{P}_{3/2}^o$ & $\mathrm{e}'$ & $J_{\mathrm{e}'} = \shrinkify{\frac{3}{2}}$ & $\omega_{\mathrm{e}',\mathrm{g}}$ \\
         ${}^2\mathrm{D}_{3/2}$ & $\mathrm{g}'$ & $J_{\mathrm{g}'} = \shrinkify{\frac{3}{2}}$ & $\omega_{\mathrm{g}',\mathrm{g}}$ \\
         ${}^2\mathrm{D}_{5/2}$ & $\mathrm{g}''$ & $J_{\mathrm{g}''} = \shrinkify{\frac{5}{2}}$ & $\omega_{\mathrm{g}'',\mathrm{g}}$ \\
          \hline \hline
    \end{tabular}
    \caption{\label{tab:Notation}Shorthand notation for the five manifolds under consideration.}
\end{table}

\begin{figure}
    \centering
    \includegraphics[width=0.95\textwidth]{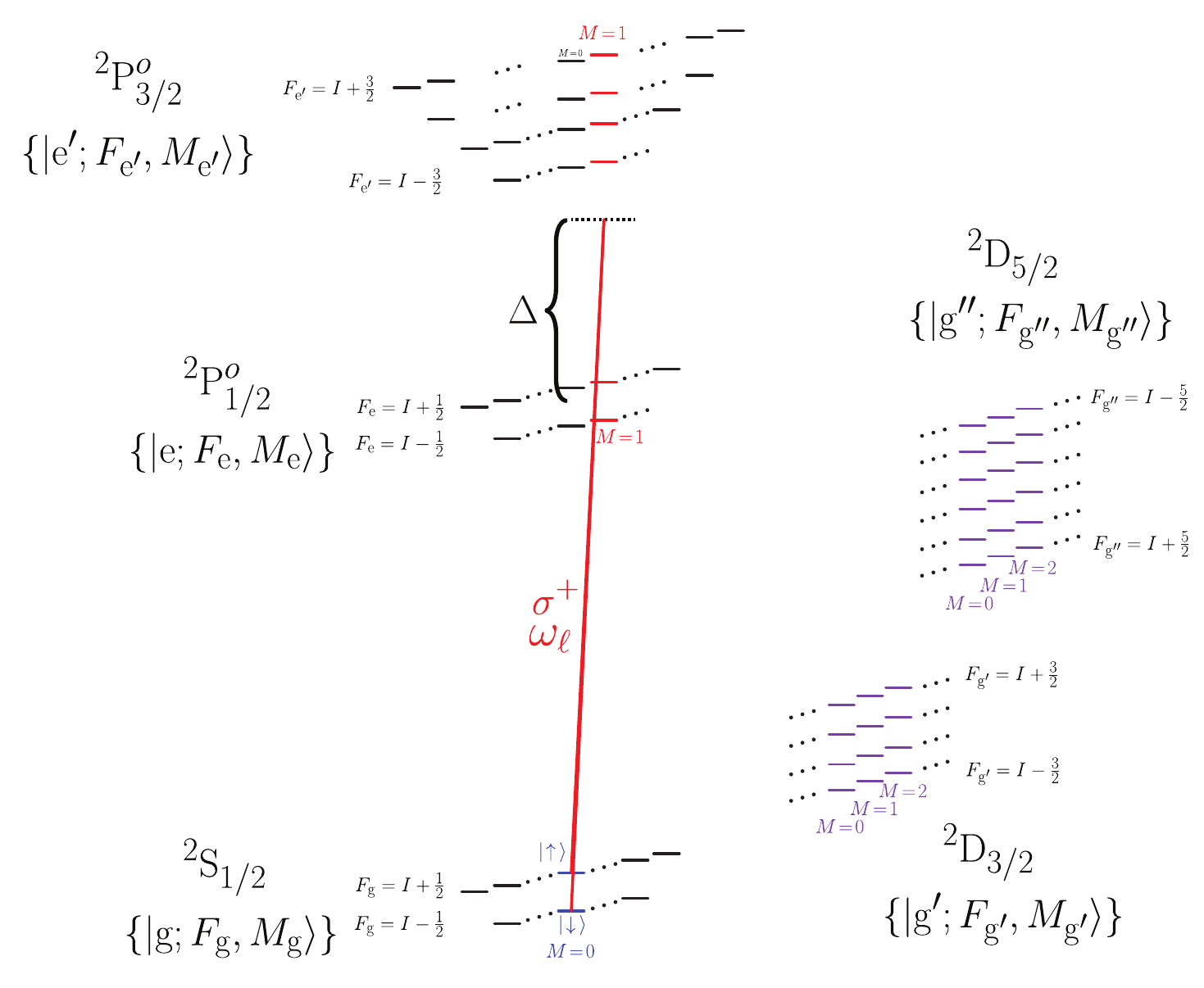}
    \caption{Level diagram relevant for driving a stimulated Raman transition between zero-field $g$-type clock states in a half-integer-spin isotope of an alkaline earth ion with pure $\sigma^+$ light.  For isotopes with $I < \frac{5}{2}$, the allowed values of $F$ are constrained to $F \ge 0$.}
    \label{fig:SILevelDiagram}
\end{figure}

The $\omega^3$-theory model, while being a fairly clear improvement over the CDA model in the regime explored experimentally in this paper, neglects a number of effects that should be considered before applying this analysis to other regimes.  As discussed in Ref.~\cite{Moore22}, there are terms arising from the counter-rotating part of the electric field (\textit{i.e.}\ emission-first processes) that can play a role, as well as coupling to more excited states than just the five manifolds considered here (Fig.~\ref{fig:SILevelDiagram}).

\end{document}